\documentclass[twocolumn,aps,pre,showpacs,amsmath,amssymb]{revtex4}
\usepackage{epsfig}
\usepackage{graphicx}
\usepackage{dcolumn}
\usepackage{bm}
\usepackage[colorlinks=true,dvipdfm]{hyperref}

\begin{document}

\title{Scaling of the entanglement spectrum in driving critical dynamics}
\author{Qijun Hu}
\author{Shuai Yin}  \email{zsuyinshuai@163.com}
\author{Fan Zhong}  \email{stszf@mail.sysu.edu.cn}
\affiliation{State Key Laboratory of Optoelectronic Materials and Technologies, School of
Physics and Engineering, Sun Yat-sen University, Guangzhou 510275, People's
Republic of China}
\date{\today}

\begin{abstract}
We present a scaling theory for the entanglement spectrum under an external driving. Based on the static scaling of the Schmidt gap and the theory of finite-time scaling, we show that the Schmidt gap can signal the critical point and be used to estimate the critical exponents no matter in the finite-size scaling region or in the finite-time scaling region. Crossover between the two regions is also demonstrated. We verify our theory using both the one-dimensional transverse-field Ising model and the one-dimensional quantum Potts model. Our results confirm that the Schmidt gap can be regarded as a supplement to the local order parameter.
\end{abstract}

\pacs{ 03.67.Mn, 64.60.De, 64.60.Ht, 64.70.Tg}
\maketitle
\section{Introduction}
Over the past decades, studies on the properties of entanglement have provided deep insight in understanding quantum matters and quantum phase transitions \cite{Osborne,JVidal,Wu,Li,Thomale}. The entanglement is usually measured by an entanglement entropy and an entanglement spectrum \cite{Osterloh,Popescu,Brandao,Amico,Eisert}. Both the entanglement entropy and the entanglement spectrum can be used to characterize the phase of quantum matters. For example, a topological entanglement entropy is proposed to identify the intrinsic topological order \cite{Preskill,Wen,Jiang}, while the degeneracy of the entanglement spectrum is used to classify the symmetry-protected topological order in one dimensional (1D) spin chains \cite{Pollmann1,Pollmann2}. Aside from the applications in characterizing the phases of matter, the scaling behavior of the entanglement \cite{Pollmann3,Fagotti,Chiara,Lepori,Rao} near a quantum critical point is proposed to locate the quantum critical point and estimate the critical exponents \cite{Zhao}. As a significant characterization in condensed matter physics, entanglement can be detected in experiments and numerical simulations \cite{Walborn}.

On the other hand, non-equilibrium dynamics of quantum many-body systems has also attracted attention \cite{Dz,Pol}. These studies are partly motivated by the progress of technologies in detecting and manipulating dynamic phenomena in cold atom experiments \cite{Greiner,Meinert}. For the driving dynamics (quench) of changing the distance $g$  to a critical point with a rate $R$, the Kibble-Zurek mechanism (KZM) divides the evolution into two adiabatic stages and an intervening impulse stage \cite{Dz,Pol,Zurek,Kibble}. In the impulse stage, according to the theory of finite-time scaling (FTS) \cite{Zhong1,Zhong2,Yin,Yin2}, the external time scale $\zeta_{\it{d}}\sim R^{-z/r}$, which is induced by the external driving, dominates the dynamics, since it is smaller than the intrinsic time scale $\zeta_{S}\sim |g|^{-\nu z}$, where $\nu$ and $z$ are critical exponents and $r=z+1/\nu$. The FTS theory further reveals that macroscopic quantities are related to the driving rate $R$ with some exponents. For example, the number of topological defects $n$, which are produced as a result of the diabatic process in the impulse stage, is scaled as $n\sim R^{-1/r}$. This is consistent with the prediction of the KZM \cite{Kibble,Zurek}.

As the entanglement provides another way to understand quantum criticality besides local operators, it is meaningful to investigate its nonequilibrium dynamic scaling. In the driving dynamics, it has been shown that the entanglement entropy is proportion to the logarithm of the driving rate at the critical point \cite{Dz,Pol,Cincio}. In real-time relaxation dynamics, the entanglement entropy increases linearly with time at the short-time stage \cite{Cala,Scha}, while in imaginary-time relaxation, it increases as a logarithm of time at the critical point \cite{Yin3}. Besides investigations on the entanglement entropy, the dynamic scaling of the entanglement spectrum has also been studied in both driving and relaxation dynamics \cite{Torlai,Canovi}. In the transverse-field Ising model, for the driving dynamics on which we shall focus in the following, it has been shown that at the critical point, the Schmidt gap, $\Delta\lambda$, defined as the difference between the two largest entanglement spectra, increases with the driving rate as a power law $\Delta\lambda\sim R^\alpha$, with $\alpha$ being estimated to be $0.064$ \cite{Torlai}. In addition, for large driving rates, no finite size effects have been observed \cite{Torlai}.

Questions arisen are then: (i) How to characterize the scaling behavior of the Schmidt gap in the whole driving process? (ii) Whether is the exponent $\alpha$, reported in Ref.~\onlinecite{Torlai}, a new exponent or a combination of other known exponents? (iii) Is there a unified scaling theory to characterize the driving dynamics for large and small driving rates? (iv) Are these results applicable in other models?

To answer these questions, in this paper, we propose a dynamic scaling theory to describe the universal dynamics of the Schmidt gap according to the theory of FTS. According to the theory, we shall show that the exponent $\alpha$ is not an independent exponent but rather is $\beta/\nu r$, where $\beta$ is the critical exponent of order parameter. The crossover behavior between the FTS region and the finite-size scaling (FSS) region is also studied. A dynamic method to determine the critical point and critical exponents is proposed. To confirm the validity and show the universality of the scaling theory, both the $1$D transverse-field Ising model and the $1$D three-state quantum Potts model are employed.

The rest of the paper is organized as follows. After introducing briefly the Schmidt gap and its static scaling behavior in Sec.~\ref{statics}, we present the dynamic scaling theory of the Schmidt gap under an external driving according to the FTS theory in Sec.~\ref{dssch}. Then, in Sec.~\ref{model} we introduce the models and our methods, while in Sec.~\ref{numr}, we verify the theory by numerical simulations of the driving critical dynamics for the models. Finally, a summary is given in Sec.~\ref{sum}.

\section{\label{statics}Brief review of the Schmidt gap and its static scaling}
\subsection{\label{schgap}Definition of the Schmidt gap}
We first introduce the definition of the Schmidt gap. An arbitrary wave function $|\psi\rangle$ can be represented into a bipartite form by the Schmidt decomposition as
\begin{equation}
|\psi\rangle=\sum_k \lambda_k |\psi^{\rm L}_k\rangle\otimes |\psi^{\rm R}_k\rangle,
\label{schdeco}
\end{equation}
where $|\psi^{\rm L}_k\rangle$ ($|\psi^{\rm R}_k\rangle$) is the $k$th eigenvector of the reduced density matrix of the left (right) subsystem and $\lambda_k$, the $k$th eigenvalue of the reduced density matrix, represents the entanglement spectrum and satisfies $\lambda_k\geq 0$ and $\sum_k \lambda_k^2=1$. $|\psi^{\rm L}_k\rangle$ and $|\psi^{\rm R}_k\rangle$ are orthogonal to each other and $\lambda_k$ is identical for the reduced density matrix of the left and right subsystems. For simplicity, we assume the array of $\lambda_k$ is arranged in a descending order with $k$. As a result, the Schmidt gap, $\Delta\lambda$, is defined as \cite{Chiara,Lepori}
\begin{equation}
\Delta\lambda\equiv \lambda_1-\lambda_2.
\label{schmidtgap}
\end{equation}
For an infinite system, which exhibits translational symmetry, $\Delta\lambda$ does not depend on the location of the lattice, while for a finite system, $\Delta\lambda$ does. Accordingly, for a finite system, we choose $\Delta\lambda$ as the one at the center point of the system. This choice can weaken the boundary effects.

\subsection{\label{staticss}Static properties of the Schmidt gap in quantum phase transitions}
It has been proposed that the Schmidt gap can be regarded as an order parameter to characterize quantum phases and quantum phase transitions~\cite{Chiara,Lepori}. To further confirm it, we compare the behavior of $\Delta\lambda$ with the local order parameter, $M$, defined, for example, as the local magnetization in the transverse-field Ising model \cite{Sachdev}. It has been shown that in the disordered phase, $M=0$ but $\Delta\lambda\neq 0$, while in the ordered phase, $M\neq 0$ but $\Delta\lambda=0$. This means that $\Delta\lambda$ is a complementary quantity to the local order parameter to describe the different phases \cite{Chiara,Lepori}.

The static scaling behavior of $\Delta\lambda$ can be obtained as follows. Note that, it has been shown that the critical exponent related to $M$ is identical with that to $\Delta\lambda$ \cite{Chiara,Lepori}. In other words, under a scale transformation with a rescaling factor $b$, $\Delta\lambda$ changes to $\Delta\lambda b^{\beta/\nu}$, which is similar to the scale transformation of $M$. Accordingly, near the critical point, the full scale transformation of $\Delta\lambda$ for a system of size $L$ with a distance $g$ to its critical point reads as \cite{Chiara,Lepori}
\begin{equation}
\Delta\lambda(g,L)=b^{-\beta/\nu}\Delta\lambda(gb^{1/\nu},Lb^{-1}).
\label{scaletran}
\end{equation}

Different scaling forms have been obtained from Eq.~(\ref{scaletran}). When $|g|^{-\nu}\ll L$, the system cannot ``feel'' its size and the finite $g$ dominates the scaling behavior. In this situation, by setting $gb^{1/\nu}=1$, one obtains the scaling form \cite{Chiara,Lepori}
\begin{equation}
\Delta\lambda(g,L)=g^{\beta}f_1(Lg^{-\nu}),
\label{scalf1}
\end{equation}
where $f_i$ is the scaling function with $i$ being an integer. Conversely, when $L\ll |g|^{-\nu}$, the correlation length is truncated by the size of the system and $L$ dominates the scaling behavior. The FSS form,
\begin{equation}
\Delta\lambda(g,L)=L^{-\beta/\nu}f_2(gL^{1/\nu}),
\label{scalf2}
\end{equation}
then follows in a similar way \cite{Chiara,Lepori}. Equations~(\ref{scalf1}) and (\ref{scalf2}) have been verified in the $1$D transverse-field Ising model \cite{Chiara,Lepori}.

\section{\label{dssch}dynamic scaling theory of the Schmidt gap in driving critical dynamics}
In this section, we develop the scaling theory describing the scaling behavior of the Schmidt gap in driving dynamics according to the theory of FTS \cite{Zhong1,Zhong2}. For $g=Rt$, the competition among the driving time scale $\zeta_{\it{d}}$ and the other two time scales $\zeta_{S}\sim g^{-\nu z}$ and $\zeta_{L}\sim L^{z}$, induced by finite $g$ and $L$, respectively, controls the driving critical dynamics near the critical point. Including $R$ in Eq.~(\ref{scaletran}), we postulate the scale transformation of the Schmidt gap as \cite{Zhong1}
\begin{equation}
\Delta\lambda(g,L,R)=b^{-\beta/\nu}\Delta\lambda(gb^{1/\nu},Lb^{-1},Rb^{r}).
\label{scaledr}
\end{equation}

Then, similar to the discussion in Sec.~\ref{staticss}, different scaling forms in different regions can be obtained by choosing suitable scale factors. By setting $Rb^{r}=1$, one obtains the scaling form of FTS \cite{Zhong1,Zhong2},
\begin{equation}
\Delta\lambda(g,L,R)=R^{\beta/\nu r}f_3(gR^{-1/\nu r},L^{-1}R^{-1/r}).
\label{scalfts}
\end{equation}
In this region, the system falls out of equilibrium and its evolution is dominated by the external driving rate, while the finite-size effects are only perturbation. In particular, in the thermodynamic limit $L\rightarrow \infty$, we obtain
\begin{equation}
\Delta\lambda(g,R)=R^{\beta/\nu r}f_4(gR^{-1/\nu r}).
\label{scalftsg}
\end{equation}

The FTS form of the Schmidt gap can also applied to determine the critical point and critical exponents similar to the order parameter \cite{Zhong1,Zhong2,Yin}. Indeed, from Eq.~(\ref{scalftsg}), one readily finds that when $\Delta\lambda(g,R)=0$, the scaled variable $gR^{1/\nu r}$ is a constant. To be explicit, for the transverse-field Ising model, $g\equiv h_{\rm{x}}-h_{\rm{xc}}$ and the transverse field values, $h_{\rm{x0}}$, at which $\Delta\lambda=0$, satisfy
\begin{equation}
h_{\rm{x0}}=h_{\rm{xc}}+cR^{1/\nu r},
\label{hx0}
\end{equation}
where $c$ is a constant. Finding $h_{x0}$ for various $R$ and fitting the results to Eq.~(\ref{hx0}), one can estimate the critical point $h_{xc}$ and the critical exponent $1/\nu r$. Then at the critical point, from Eq.~(\ref{scalftsg}), $\Delta\lambda\propto R^{\beta/\nu r}$. Accordingly, we can determine the critical exponent $\beta/\nu r$.

Similarly, by setting $Lb^{-1}=1$, Eq.~(\ref{scaledr}) leads to the FSS form \cite{Liu,Huang},
\begin{equation}
\Delta\lambda(g,L,R)=L^{-\beta/\nu}f_5(gL^{1/\nu},RL^{r}),
\label{scalfss}
\end{equation}
which returns to Eq.~(\ref{scalf2}) for $R=0$.

Note that both $f_3$ and $f_5$ can describe the scaling behavior in the FTS and the FSS region. As a result, $f_3$ and $f_5$ fulfill
\begin{equation}
f_5(X,Y)=Y^{\beta/\nu r}f_3(XY^{-1/\nu r},Y^{-1/r}),
\label{f3f4}
\end{equation}
or conversely,
\begin{equation}
f_3(X,Y)=Y^{\beta/\nu}f_5(XY^{-1/\nu},Y^{-r}).
\label{f4f3}
\end{equation}
So, at the critical point, the crossover between the FTS region and FSS region can be identical. When $L^{-1}R^{-1/r}\ll 1$, the system is in the FTS region, and $\Delta\lambda$ behaves as \cite{Zhong1,Zhong2,Liu,Huang}
\begin{equation}
\Delta\lambda(L,R)=R^{\beta/\nu r}f_6(L^{-1}R^{-1/r}),
\label{scalfts1}
\end{equation}
according to Eq.~(\ref{scalfts}) where $f_6(Y)=f_3(0,Y)$. Equation~(\ref{scalfts1}) shows that $\Delta\lambda(L,R)\propto R^{\beta/\nu r}$ when $L^{-1}R^{-1/r}$ can be neglected. Comparing the scaling behavior in Ref.~\onlinecite{Torlai}, in which $\Delta\lambda$ is proportional to $R^\alpha$ for large sizes, we find thus $\alpha=\beta/\nu r$. In contrast, when $L^rR\ll 1$, the system is in the FSS region, and $\Delta\lambda$ satisfies
\begin{equation}
\Delta\lambda(L,R)=L^{-\beta/\nu}f_7(L^{r}R),
\label{scalfts2}
\end{equation}
and $\Delta\lambda\propto L^{-\beta/\nu}$ in this region when $L^{r}R$ can be neglected, where $f_7(Y) = f_5(0,Y)$.
Moreover, according to Eq.~(\ref{f3f4}), in the FSS form, the FTS region behaves distinctly from the FSS region, because asymptotically, $f_7$ tends to \cite{Huang}
\begin{equation}
f_7(L^{r}R)\sim (L^{r}R)^{\beta/\nu r}.
\label{f6fss}
\end{equation}
On the other hand, in the FTS form, the FSS region behaves asymptotically as
\begin{equation}
f_6(L^{-1}R^{-1/r})\sim (L^{-1}R^{-1/r})^{\beta/\nu},
\label{f5fts}
\end{equation}
according to Eq.~(\ref{f4f3}) \cite{Huang}. These results will be born out by simulations shown in Figs.~\ref{ftsfss} and \ref{fssfts} for the Ising model and Fig.~\ref{pottsfssfts} for the Potts model below.
\section{\label{model}Model and numerical method}
The Hamiltonian of the transverse-field Ising model is \cite{Sachdev}
\begin{equation}
\mathcal{H}_{\rm I}=-\sum_{i=1}^{L-1} \sigma_{i}^z\sigma_{i+1}^z-h_{\rm{x}}\sum_{i=1}^L\sigma_i^x,
\label{modelI}
\end{equation}
where $\sigma_{i}^x$ and $\sigma_{i}^x$ are the Pauli matrix in the $x$ and $z$ directions, respectively, at site $i$. The critical point of model (\ref{modelI}) is $h_{\rm x}=h_{\rm xc}=1$ \cite{Sachdev}. The exact critical exponents are $\beta=1/8$, $\nu=1$, and $z=1$ \cite{Sachdev}. This model has been realized in CoNb$_2$O$_6$ \cite{Coldea}.

In order to confirm the universality of the scaling properties of the entanglement spectrum, we also employ the three-state quantum Potts model \cite{Solyom} with the Hamiltonian
\begin{equation}
\mathcal{H}_{\rm P}=-\sum_{i=1}^{L-1}\sum_{q=1}^2M^{z,q}_i M^{z,3-q}_{i+1}-h_{\rm{x}}\sum_{i=1}^{L}M^x_{i},
\label{modelP}
\end{equation}
where $M^{x}$ and $M^{z,q}$ are the three Potts spin matrices
\[
M^x=\left[
\begin{array}{ccc}
 2 & 0 & 0 \\
 0 & -1 & 0 \\
 0 & 0 & -1
\end{array}
\right],
M^{z,1}=\left[
\begin{array}{ccc}
 0 & 1 & 0 \\
 0 & 0 & 1 \\
 1 & 0 & 0
\end{array}
\right],\]
and $M^{z,2}=(M^{z,1})^2$. The critical point of model (\ref{modelP}) is again $h_{\rm x}=h_{\rm xc}=1$, and the conjectured critical exponents are $\beta=1/9$, $\nu=5/6$, and $z=1$ \cite{Solyom}.

We use the time-evolving block decimation (TEBD) algorithm \cite{Vidal} to calculate the evolution of the Schmidt gap under an external driving. After expanding the wave function into a matrix product form via the Vidal decomposition \cite{Hatano}, each site is attached by a matrix. This matrix is updated by acting the local evolution operator, which is the Suzuki-Trotter decomposition of ${\rm exp}(-i\mathcal{H}t)$. By taking the translational symmetry into account, a variant of the TEBD algotithm has been developed to calculate the dynamics of a quantum chain with infinite length. This method is called infinite time-evolving block decimation (ITEBD) algorithm \cite{Vidal1}. We choose the time interval to be $0.01$ and the number of the kept states to be $50$. Three decimal places are kept in our results. More accurate results are expected by choosing larger numbers of truncation and smaller time intervals.

We start with the ground state of the Hamiltonian for a sufficiently large $g$ and decrease it linearly with time with a given $R$. This is just the reverse of the parallel study of the order parameter, in which $g$ is increased linearly \cite{Yin}, in accordance with their complementary character.

\section{\label{numr}Verification of the dynamic scaling}
\subsection{Dynamic scaling of the Schmidt gap in the quantum Ising model}
\begin{figure}
  \centerline{\epsfig{file=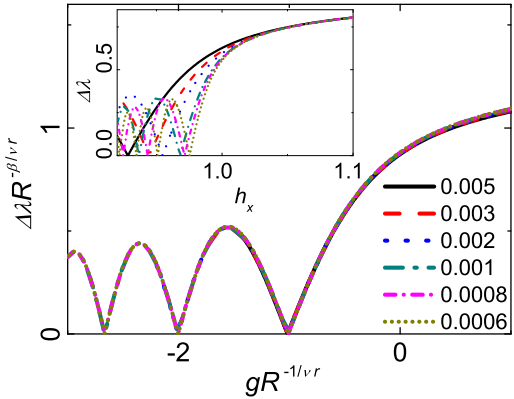,width=1.0\columnwidth}}
  \caption{\label{Isinginf} (Color online) Curves of $\Delta\lambda$ versus $R$ collapse onto each other for model (\ref{modelI}). The corresponding curves before rescaling are plotted in the inset.}
\end{figure}

In this section, we shall verify the FTS forms of the Schmidt gap for the $1$D transverse-field Ising model. We firstly verify the full scaling form of FTS in Eq.~(\ref{scalftsg}) for the infinite size situation. Figure~\ref{Isinginf} shows the evolutions of the Schmidt gap for various driving rates. Near the critical point, the curves collapse onto each other after rescaling with the exact critical point and critical exponents as input. This result confirms the FTS form of Eq.~(\ref{scalftsg}) and shows that $\Delta\lambda$ has the same scaling dimension as the local order parameter $M$ even in the nonequilibrium situation.

\begin{figure}
  \centerline{\epsfig{file=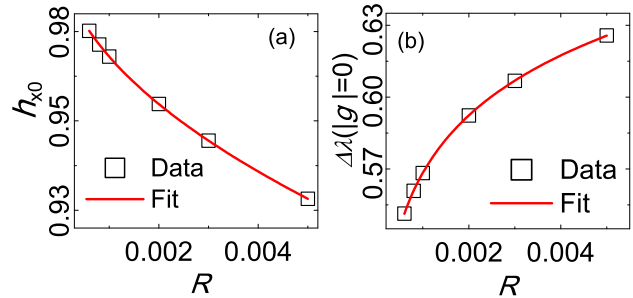,width=1.0\columnwidth}}
  \caption{\label{dcpce} (Color online) Estimation of the critical point and critical exponents for model (\ref{modelI}). (a) Fitting of $h_{\rm{x0}}$ for the critical point $h_{\rm{xc}}$ and the critical exponent $1/\nu r$ according to Eq.~(\ref{hx0}). (b) Fitting of $\Delta\lambda(|g|=0)$ for the critical exponent $\beta/\nu r$ according to Eq.~(\ref{scalftsg}).}
\end{figure}

Figure~\ref{dcpce} shows the procedure to determine the critical properties according to Eq.~(\ref{hx0}). The critical point $h_{\rm{xc}}$ is estimated to be $h_{\rm{xc}}=1.000$, in good agreement with the exact value, with the critical exponents $1/\nu r=0.506$ and $\beta/\nu r=0.060$. By substituting $z=1$, these exponents lead to $\beta=0.118$ and $\nu=0.976$, which are close to the exact values $\beta=1/8$ and $\nu=1$.

\begin{figure}
  \centerline{\epsfig{file=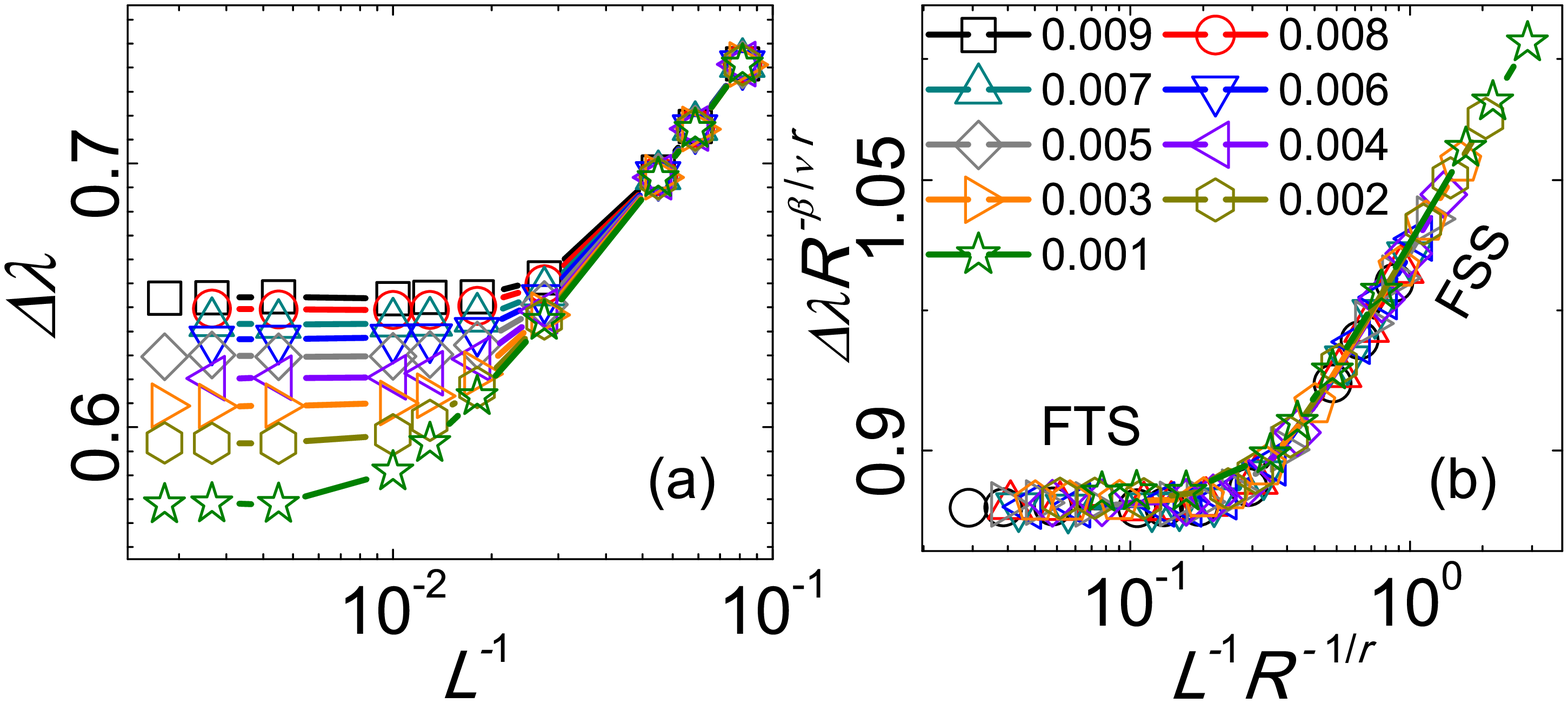,width=1.0\columnwidth}}
  \caption{\label{ftsfss} (Color online) (a) $\Delta\lambda$ versus $L^{-1}$ for various $R$ at the critical point of model (\ref{modelI}). (b) $\Delta\lambda R^{-\beta/\nu r}$ versus $L^{-1}R^{-1/\nu r}$ according to Eq.~(\ref{scalfts1}). The two different regions of FSS and FTS are marked. Lines connecting symbols are only a guide to the eye. Double logarithmic scales are used.}
\end{figure}

Then we focus on the the finite-size effects and verify Eq.~(\ref{scalfts1}). At the critical point, the dependence of $\Delta\lambda$ on $L^{-1}$ for various $R$ is shown in Fig.~\ref{ftsfss}(a). From Fig.~\ref{ftsfss}(a), we see that for large $L$, the curves are independent of $L$. In contrast, for small $L$, the curves for different driving rates collapse onto each other and are independent of the driving rate. This is because in the former case $L$ is bigger than $\xi_{\it{d}}$ and FTS dominates, whereas in the latter case, their relation is reversed, and the evolution is dominated by the finite-size effects.

Figure~\ref{ftsfss}(b) shows that the rescaled curves collapse onto each other, confirming Eq.~(\ref{scalfts1}). From Fig.~\ref{ftsfss}(b), we find that when $L^{-1}R^{-1/r}$ is small, the rescaled value, $\Delta\lambda R^{-\beta/\nu r}$ is almost a constant. This constant is just $f_5(0)$ in Eq.~(\ref{scalfts1}). For large $L^{-1}R^{-1/r}$, the system enters the FSS region, in which the curve is almost a straight line with a slope of $0.124$ in the double-logarithm scales. This confirms Eq.~(\ref{f5fts}), which predicts the power-law relation in the FSS region with an exponent of $\beta/\nu$ \cite{Liu,Huang}, whose exact value is $0.125$.

\begin{figure}
  \centerline{\epsfig{file=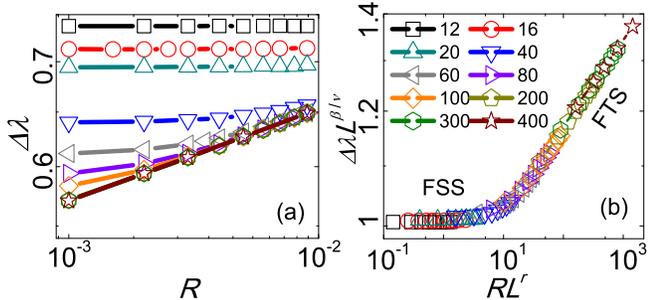,width=1.0\columnwidth}}
  \caption{\label{fssfts} (Color online) (a) $\Delta\lambda$ versus $R$ for various $L$ at the critical point of model (\ref{modelI}). (b) $\Delta\lambda L^{\beta/\nu}$ versus $R L^{r}$ according to Eq.~(\ref{scalfts2}). The two different regions of FSS and FTS are marked. Lines connecting symbols are only a guide to the eye. Double logarithmic scales are used.}
\end{figure}

We can also study the crossover from the FSS form according to Eq.~(\ref{scalfts2}). Figure~\ref{fssfts}(a) shows that for small $L$, the system is in the FSS region, in which $\Delta\lambda$ is almost a constant; while for large $L$, the system is in the FTS region and the curves collapse onto each other again. Figure~\ref{fssfts}(b) shows that the rescaled curves collapse onto each other and confirms Eq.~(\ref{scalfts2}). In the FTS region, power function fitting gives the exponent to be $0.062$, which is close to the exact value $\beta/\nu r=0.0625$. This result confirms Eq.~(\ref{f6fss}). Note that this is also the slope of the asymptotic straight line for large $L$ in Fig.~\ref{fssfts}(a) and is thus $\alpha$ in Ref.~\onlinecite{Torlai}. In fact, Fig.~\ref{fssfts}(a) is just Fig. 8 of Ref.~\onlinecite{Torlai}, albeit with smaller $L$ and $R$. Therefore, our scaling theory explains well the results of Ref.~\onlinecite{Torlai}.

\subsection{Dynamic scaling of the Schmidt gap in the quantum Potts model}
\begin{figure}
  \centerline{\epsfig{file=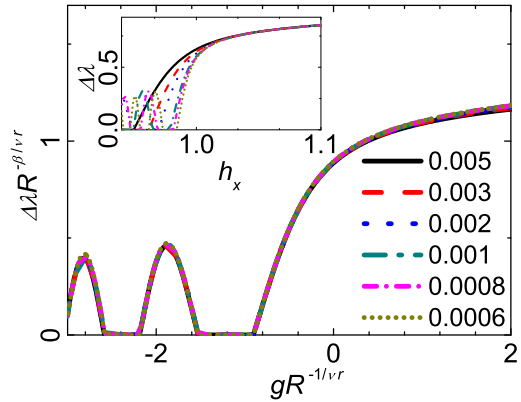,width=1.0\columnwidth}}
  \caption{\label{pottsfts} (Color online) $\Delta\lambda$ versus $R$ collapse onto each other for model (\ref{modelP}). The corresponding curves before rescaling are plotted in the inset.}
\end{figure}

To further confirm the scaling theory, we study the driving dynamics of the quantum Potts model (\ref{modelP}). First, we consider the infinite-size situation. Figure~\ref{pottsfts} shows that the curves for different driving rates collapse onto each other after rescaling by substituting the critical exponents for the quantum Potts model. This result confirms Eq.~(\ref{scalftsg}) and demonstrates the universality of Eq.~(\ref{scalftsg}) in the driving dynamics. Comparing Fig.~\ref{pottsfts} with Fig.~\ref{Isinginf}, one finds that in Fig.~\ref{Isinginf}, the Schmidt gap has some discrete points in which $\Delta\lambda=0$, while in Fig.~\ref{pottsfts}, there is a continuous region corresponding to $\Delta\lambda=0$. The reason for this difference may be the symmetry group for the quantum Potts model is $Z_3$ while for the Ising model is $Z_2$. In spite of this difference, the scaling theory can well describe the driving dynamics for both models.

\begin{figure}
  \centerline{\epsfig{file=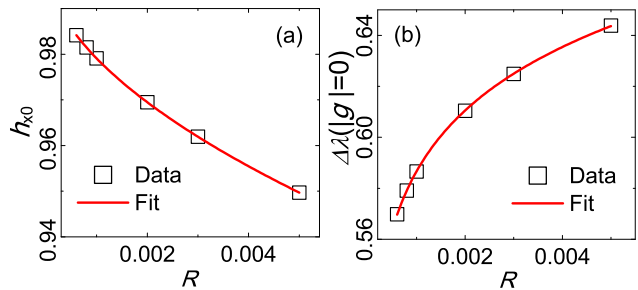,width=1.0\columnwidth}}
  \caption{\label{dpcpce} (Color online) Estimation of the critical point and critical exponents for model (\ref{modelP}). (a) Fitting of $h_{\rm{x0}}$ for the critical point $h_{\rm{xc}}$ and the critical exponent $1/\nu r$ according to Eq.~(\ref{hx0}). (b) Fitting of $\Delta\lambda(|g|=0)$ for the critical exponent $\beta/\nu r$ according to Eq.~(\ref{scalftsg}).}
\end{figure}

Similar to the situation in the quantum Ising model, we can also use the method according to Eq.~(\ref{scalftsg}) to determine the critical point and the critical exponents. The only difference is that we should choose the points just at which $\Delta\lambda$ becomes zero as the ``characteristic" points to determine the critical point, since the zeros for the Potts model are continuously distributed. Figure~\ref{dpcpce} gives the fitting results. The critical point is estimated to be $h_{\rm{xc}}=1.000$ with the critical exponents $1/\nu z=0.546$ and $\beta/\nu r=0.0576$. These exponents result in $\beta=0.106$ and $\nu=0.832$ by substituting $z=1$. They are all close to their exact and conjectured values.

\begin{figure}
  \centerline{\epsfig{file=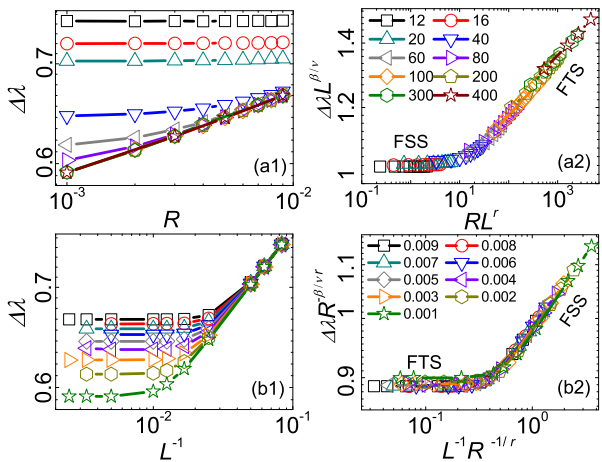,width=1.0\columnwidth}}
  \caption{\label{pottsfssfts} (Color online) FSS and FTS and their crossover of the Schmidt gap for the Potts model (\ref{modelP}). (a1) Dependence of $\Delta\lambda$ on $R$ for various $L$ given in (a2). (a2) The rescaled curves from (a1) according to Eq.~(\ref{scalfts2}). (b1) Dependence of $\Delta\lambda$ on $L^{-1}$ for various $R$ given in (b2). (b2) The rescaled data from (b1) according to Eq.~(\ref{scalfts1}). Lines connecting symbols are only a guide to the eye.}
\end{figure}

Next, we study the finite-size effects in the driving dynamics near the critical point. Figure~\ref{pottsfssfts}(a1) shows the dependence of $\Delta\lambda$ on $R$ for various $L$. The rescaling curves, obtained by rescaling the data with the size $L$, collapse onto each other well, as displayed in Fig.~\ref{pottsfssfts}(a2). Similar to the situation in the transverse-field Ising model, for small $RL^r$, the system is dominated by the finite-size effects, and $\Delta\lambda L^{\beta/\nu}$ tends to a constant, indicating that $\Delta\lambda\propto L^{-\beta/\nu}$. For large $RL^r$, the system is in the FTS region. According to Eq.~(\ref{f6fss}), the slope is $\beta/\nu r$. This exponent is fitted to be $0.060$, which is close to the exact value $0.061$.

We can also present the above results in the FTS form. Figure~\ref{pottsfssfts}(b1) shows the dependence of $\Delta\lambda$ on $L$ for various $R$, while Fig.~\ref{pottsfssfts}(b2) displays the collapses after rescaling. The slope of the FSS region is estimated to be $0.112$, which is close to the conjected value of $\beta/\nu$. These results confirm the scaling analysis in Eq.~(\ref{f5fts}).

\section{\label{sum}Summary}
We have studied the scaling behavior of the entanglement spectrum in the driving dynamics. A scaling theory based on finite-time scaling has been developed to describe the scaling behavior of the Schmidt gap of a system under an external driving. We have found that the Schmidt gap develops similar scaling behavior as the local order parameter does. This shows that the Schmidt gap can be regarded as a supplement to the local order parameter. To illustrate it, we have utilized the FTS form of the Schmidt gap to determine the critical point and critical exponents. The crossover behavior between the FSS region and the FTS region has also been explored and the previously published results have been well explained within the present theory. The universality of the scaling theory are confirmed in both the transverse-field Ising model and the three-state Potts chain.

\end{document}